# Quantum gauge confinement of multiple quarks based on the homogeneous 5D projection theory

Kai-Wai Wong[1], Gisela A. M. Dreschhoff[1], Högne J. N. Jungner[2]

[1]Department of Physics and Astronomy, University of Kansas, Lawrence, Kansas, USA, [2]University of Helsinki, Helsinki, Finland

**Abstract**

A quick and simplified review of the 5D quantum field theory is presented. The role of topological mapping, which must preserve gauge invariance, is done in two ways, leading to the realization of the gauge transformation in the 5D space-time becoming two separate gauge constraints, one for the multi-quark state quark constituents, while the other is the quantum confinement imposed on the gluon potentials, formed from products of vector potentials generated by products of the fractional charged quark currents. The procedure presented clearly shows multi-quark states can be designed and that they can be verified by experiments, such as the penta-quark state reported. Based on these gauge constraints we propose the existence of 4, 5 and 6 quark states.

## 1. Introduction

Recently there has been some excitement about the confirmed finding of a penta-quark state [1]. In fact we did propose another such state formed of $\pi^+$ and proton [2]. While previously, based on old understanding of the standard model [3], it was believed that such penta-quark states are unlikely [4]. In addition we had also interpreted the p-p state of 125 GeV as that of a hexa-quark state, and not that of the Higgs boson [5]. Because of these rather controversial results, we believe a short note clearing the quantum gauge confinement on multiple quark states based on the homogeneous 5D projection theory into the SU(3)xL is called for. Where L is the 4D Lorentz space-time, and the generators of the SU(3) Lie group are the up, down, and top quarks [6]. The remaining bottom, charm and strange quarks, although having the same 2/3 and 1/3 e charges, are superpositions of the generator set [7].

To illustrate the gauge confinement, we need to review briefly why a quantum theory derived from the 5D homogeneous space-time implies gauge invariance, and any topological mapping from 5D spinors to 4D space-time spinors must preserve the gauge invariance. It was shown in ref. [6] that the 5D homogeneous quantum operator solutions are massless fields, represented by 5 component vector potential fields, and from the linearized form of the operator, the solutions are massless spinor fields, which will couple to the vector fields, with the coupling constant denoted by the positive and negative unit charge e, namely e-trinos and anti-e-trinos. Hence, it is obvious that the decoupling of these vector and spinor solutions is the gauge transformation, leading to the quantization of flux, as n(h/e), where n is an integer. Because we start with a homogeneous 5D manifold, the creation of e-trino, anti-e-trino pairs must conserve zero net charge and momentum, thus if the pair state forms a closed gauge loop, with open boundary it is in the form of a letter 8, with one circulating circle loop that of the e-trino, and the other opposite circulating circle loop that from the anti-e-trino. However, if they are confined by a closed spherical boundary, then the two circle loops are concentrically aligned.

Each circle loop then obeys the flux quantum n(h/e), where n is an integer, n=1,2…, making the total flux within the letter 8 loop, (n+n')h/e, not n(h/2e) as in an off-diagonal-long-range-order pair Cooper state for a single charged spinor. The reason is that for a Cooper pair, the loop circulation must be for the pair center of mass. As mentioned earlier, this therefore gives the lowest gauge loop for a pair of oppositely charged massless spinors equal to 2h/e.

In order to get from 5D to 4D, a projection from one of the 4 space component onto the remaining time and 3D space components must be performed, leading to the Poincaré mapping, and thus producing a boundary given by SU(2)×SU(3)×L, enclosing a 4D space, time independent void. In order to preserve the gauge transformation for either charged massless spinor, the time projection $P_0$ leads to a SU(2) symmetry, with solutions given by the SU(2) generators, which are the 3 leptons together with the charge-less massless neutrinos, as one set of the charge is eliminated by time irreversibility, a property of the homogeneous 5D metric. On the other hand, the conformal projection $P_1$ from one space component onto the remaining 3 components, with geometric symmetry preserved for the remaining 3 components, results in a set of $s_i$ fractional charges, given by $s_i e$, where $s_i$ is equal to (2/3), (-1/3) and (2/3) thus forming the SU(3) generators. Such a symmetric group applied to the gauge transformation requires an invariant constant $\alpha = |s_i| e / m_i$, where $m_i$ is the $P_1$ component projected onto the resulting 3 momenta [6]. However, each of these remaining 3 component spinor states do not preserve the original gauge invariance as they couple to the 4D massless Maxwell vector potentials. For the 5D massless vector potentials it is easy to see that a reduction in the 4 space coordinates to 3 space-components, while preserving space homogeneity, only split into 2 parity Maxwell potentials. Thus to preserve the gauge invariance, we need first to construct products of the fractional charge spinors, such that their charge sum remains integer multiple of the unit e. It is due to this principle that we get the Gell-Mann standard model. Because of the fact that an in phase circulating loops state formed by the two oppositely charged massless spinors, having no net charge, but must have an angular momentum, which must be cancelled by that provided from the boundary condition which is now possible as the boundary contains finite mass, the gauge constraint is now extended to n(h/e), with n including the integer 0.

However, this alone is insufficient, as each of these fractional charged spinors form charged currents, and through Maxwell equations generate gluon fields by products of these multiple quark currents. Such products must also satisfy gauge constraint. Because of the charged fractions, the gauge constraint, which now extends to n=0, 1, 2, …, requires at least the sum of 2 fractions, which creates a Boson field that we identify as mesons. To obtain the same constraint for a Fermion field, we require at a minimum the product of 3 fractional charged spinors, thus giving us the gluon fields for the baryons.

The extended discussion is found in the following two sections.

## 2. Meson gluon field, and meaning of gauge confinement

To illustrate how gauge confinement plays the roll in multi-quark states, we shall use the meson as an example. Its mass, as well as its quark constituent signature are the result of applying gauge invariance from the conformal projection of the 4th space component onto the remaining 3 space components within the Lorentz manifold. In Chapter 7, of ref. [6] we showed explicitly the conformal projection $P_1$ acting on the spinor equation, and considering its coupling to the Maxwell vector potential we obtain the solution split into a set of quarks, with fractional charges (2/3)e, (-1/3)e and (2/3)e, which corresponds to the u, d, t quarks, represented by eq.(7.8) in ref. [6]:

$$\Psi_i = e^{i\alpha m s_i \Phi_0} \times \Psi \quad (2.1)$$

α is the invariant constant ($s_i \times e/s_i \times m$), and $s_i$ is the quark charge fraction 2/3, -1/3 and 2/3, while m is the bare quark mass, $\Phi_0$ is the quantum flux h/e, and Ψ is a 4D space-time quark spinor, solution of the linear operator:

$$P_1\{\tilde{\gamma}^\mu \widetilde{\partial_\mu}\}P_1^* \qquad (2.2)$$

where μ runs from 0 to 4. The ~ sign on top, implies 5D. See chapter 7 in ref. [6] for details.

It is clear, because $s_i$ is a fraction, $\Psi_i$, given by eq.(2.1) is not gauge invariant, as we can express $\Phi_0$ as the closed loop integral of the Maxwell vector potential. It is this condition that leads to the gauge constraint for hadrons, as products of multiple $\Psi_i$.

For mesons, it is the product of two quark spinors $s_i$ and $s_j$, such that

$$s_i + s_j = n \qquad (2.3)$$

where n is an integer, including n=0. Thus $s_i \times s_j$ represent the meson signature, and has no relationship to the gluon potential of the meson, which is generated by quark currents on the meson surface. These currents are given by $s_i e v_\mu \psi_i \psi_i^*$, which actually are created from the vacuum, and do not have a net particle number. For the mesons, however, such currents must be created in pairs, and due to gauge constraint must satisfy $s_i+s_j=n$ of eq. (2.3). This pair is not the same as the meson signature pairs, which are stationary within the meson structure. Hence the gluon potential generated is totally enclosed by these surface quark currents. To express the gauge confinement means that the enclosed gluon potential must be expressed by the product of the two vector potential fields, given by a tensor $\Gamma_{ij}$, which is a constant evaluated at the radius value of the meson structure, which is considered to be spherical. This result is the result of Perelman mapping [9].

As discussed, the gluon potential for mesons is due to the product of two Maxwell potentials generated by a pair of currents of fractional charges and not due to those quarks within the meson which do not give rise to charged currents, as given by eq. (8.7) in ref. [6]. Therefore to preserve gauge invariance, required by the 5D fields, irrespective of resulting projections, the gauge condition is split into 2, one dealing with the meson quark signature, and the second dealing with the evaluation of the gluon potential value. The mathematics for the gauge transformation due to dimension reduction is a very involved problem of topologic mapping, as shown by Perelman mapping on proving the proof of the Poincaré Conjecture [8, 9], or the Chern-Simons gauge transformation derived for the 2D hydrogen problem [10, 11]. It is not our intension to show the rigorous proof here. Rather we like to present a simple procedure, such that we can design multi-quark states that can be verified from experiments, such as the penta-quark state reported [1].

All gluon potentials from products of multiple quark currents are positive. As such a meson gluon field is generated by a pair of quark currents, and it is coupled to the quarks within the meson, and the product of such currents must also satisfy the gauge constraint. In fact, here, these currents must form a closed loop in order to generate the gluon field. Because this pair satisfies the quantization of flux, the lowest flux is given by h/Q, where Q is irrespective of the charge of its current. Hence if this current generates a vector potential $A_\mu$, then this $A_\mu$ satisfies:

$$\oint A_\mu dx_\mu = 2\pi r A_\mu = h/Q \qquad (2.4)$$

where r is the loop radius perpendicular to the loop circumference and hence to $A_\mu$.

Since $\Gamma_{ij}$ is a function of r, then $A_\mu$ generated by the $\Psi$ quark current must also be a function of r. It is this feature that implies that for mesons the current generated by the pair is confined by the flux quantum h/e, irrespective of its signature in terms of its charge. As such the gluon potential that is coupled to the meson field, must be evaluated at the simple circular radius of the flux quantum h/e, totally irrespective of the net charge signature. For example: All three different charged pion states have the exact same gluon potential value [12]. A similar argument applies to the baryon gluon potential given by the product of 3 vector potentials. Readers are advised to read ref. [6].

## 3. Topological twisting of the gauge loop for the 4, 5, and 6 quark states

Because of the charge fractions (2/3)e and (-1/3)e character of quarks, any number of such quark products greater or equal to 4 can be represented by direct products of two gauge loops, each satisfying n(h/e). Hence for a single simple circular loop that confines all the quarks in a multi-quark current state having 4, 5 and 6 quarks, must at least satisfy 2(h/e), the excited flux quantum. The transformation from one circular loop into a product of a quantum of two (h/e) is allowed topologically when n for the single circle loop is 2, when the single loop is twisted into the letter 8. Such a topological transformation reduces the gluon potential and will result in the doubling of the individual letter 8 circle radius as the gluon potential value is decreased. This transformation however is allowed only if the boundary is open. The vector potential generated by a charge e, having velocity v, is inversely proportion to $r^2$. Thus if we integrate over a closed loop of radius 2r, will reduce the flux by 1/2, as compared to the original lowest flux quantum 2(h/e), hence is of value h/e. which gives the sum of the two circle loops, and then conserve the original flux value of 2(h/e). It is such a 2(h/e) single circle loop excited flux quantum that allows for the higher number of quark states to be quantum gauge confined. However, that does not mean necessarily it can be transformed by twisting into the letter 8 loop, to give a lower mass energy state. To show the lower energy state existence, it must possess a net attractive binding Coulomb potential between the two circle loops that gauge confine the multiple quarks within each part of the high number quark state. As an explicit example, we refer to the proton-proton 125 GeV state published earlier [5].

This 6 quark state is the product of 4 u, and 2 d quarks. Its net charge is 2e. Thus the only bound state is the single circular gauge loop, which gives the flux quantum of 2(h/e), as when mapped into the letter 8 loops, each half is repulsive towards the other. Hence when this loop is twisted into the letter 8, then the gluon potential is reduced by a factor of $(2)^6$! and it is down to just 2 GeV and will decompose into two unbounded protons. Thus this state is not stable. Furthermore, before the breakup, the inter quark Coulomb potential for the proton formed by equilateral geometry on the distances between any pairs, gives a net Coulomb potential zero [12], and thus there is no force to bind the two halves. In the letter 8 gauge loop we will not have a net quantum well to make this 6 quark state with lower total mass than that of two unbounded protons. Hence no such state can be observed experimentally.

On the other hand, let us consider another 6 quark state that is the combination of a proton and a neutron. It is represented by 3u, and 3d quarks. Its net charge is now only e. A single gauge circular loop for the gluon potential is still given by n(h/e), where n=1, 2, ... This loop twisted into the letter 8 separates the gluon potential into direct sums of two, roughly equal to 2 times the proton mass as in the case for the p-p state, except the neutron has a net attractive Coulomb potential equal to $-1/3(e^2/r)$, where r is equal to the diameter of each circle of the letter 8 loop, and is of the magnitude of the proton size. This spherically symmetric negative potential due to quantum theory must be treated as equally divided between both circles, and not just for the neutron, as the neutron can be in either of the letter 8 circles. Thus it must be treated as a quantum well, where protons and neutrons, without corrections for the inter-quark electromagnetic contributions, are placed. The eigenstates of the proton and neutron then form the inter nucleon binding according to shell model theory [13]. Hence this state will have a smaller mass energy than the sum of a free proton and a free neutron. It is this very fact that nuclei with different proton and neutron numbers can be energetically formed, and for stable heavy nucleons, the number of neutrons exceeds the number of protons.

For the penta-quark states, it is the product of a meson and a baryon. The meson quark state is a two body problem. Its Coulomb potential is repulsive for all charged states, while all neutral mesons have attractive Coulomb potential. Thus if a penta-quark state includes a 3 quark proton, then the meson Coulomb potential determines whether the letter 8 gauge loop gives a stable lowest penta-quark state. In short only the product between a neutral meson and the proton would have such a lowest stable state, like the J/Ψ-p state and is observable, as confirmed by the CERN result from a product of the J/Ψ and proton. [1].
The two charged pions with the proton will have a mass of roughly 33.8 GeV [2]. The difference between $\pi^+$ and $\pi^-$ is that the penta-quark state, when split into $\pi^+$ and proton will not be able to form an atomic-like bound state, while for $\pi^-$ it will form such an atomic bound state with the proton. These states must have a Bohr radius greater than the proton size, unless this hydrogen-like state is confined in 2 dimensions, then the Bohr radius could be shrunk to the surface of the proton, leading to a collapse, with its mass given by the sum of the pion and proton plus the net kinetic energy of the reduced mass, which is much less than that of 33.8 GeV for the $\pi^+$ and proton bound state [2].

## 4. Summary

A quick and simplified review on the 5D quantum field theory is presented. The projection leading to the reduction from 4 space components to 3 space components is divided into a time projection, and a conformal projection. The time projection gives the formation of a SU(2)×L manifold, where L is the Lorentz space-time, and the generators for the SU(2) Lie group are the 3 pairs of leptons with their respective neutrinos, while the conformal projection from one space coordinate onto the remaining 3 coordinates, leads to the SU(3)×L manifold. The quantum spinors then form the Gell-Mann standard model, build from 3 pairs of quarks: u,d; t,b; and c,s; with fractional charges of (2/3)e and (-1/3)e. It is shown that the gauge transformation in the 5D space-time becomes two separate gauge constraints, one for the multi-quark states quark constituents, while the other is the quantum confinement imposed on the gluon potentials, formed from products of vector potentials generated by products of the fractional charged quark currents. It is based on these gauge constraints, that we propose the existence of 4, 5 and 6 quark states. Based on the model discussed, it should be possible to experimentally find more, not yet observed multi-quark states.